\documentclass[prd, superscriptaddress, tightenlines, showpacs, showkeys, aps, amsfonts, amssymb, twocolumn,
reprint, nofootinbib,preprintnumbers]{revtex4}
\usepackage{amsmath}
\usepackage{graphicx,epsfig,float,color}
\usepackage[caption=false]{subfig}
\usepackage{shorthand,hyperref,enumitem}
\hypersetup{colorlinks = true}
 \usepackage[update,prepend]{epstopdf}

\begin{document}

\title{Coupling Extraction From Off-Shell Cross-sections}

\author{Baradhwaj Coleppa} \email{baradhwa@email.arizona.edu}
\affiliation{Department of Physics, University of Arizona, Tucson AZ 85721, USA}
\author{Tanumoy Mandal} \email{tanumoymandal@hri.res.in}
\affiliation{Regional Centre for Accelerator-based Particle Physics, Harish-Chandra Research Institute, Chhatnag Road, Jhusi, Allahabad - 211 019, India.}
\author{Subhadip Mitra} \email{subhadip.mitra@th.u-psud.fr}
\affiliation{Laboratoire de Physique Th\'{e}orique, CNRS - UMR 8627, Univ. Paris-Sud 11, F-91405 Orsay Cedex, France}

\date{\today}

\begin{abstract}
In this note, we present a novel method of extracting the couplings of a heavy particle to the Standard Model states. Contrary to the usual discovery process which involves studying the on-shell production, we look at regions away from resonance to take advantage of the simple scaling of the cross-section with the couplings. We illustrate the procedure with the example of a heavy quark.
\end{abstract}


\maketitle

\section{Introduction}

With the discovery of the Higgs boson \cite{HiggsDisco}, the particle content of the Standard Model (SM) is well established. The future aim of the Large Hadron Collider (LHC) lies in uncovering clues for Beyond the SM (BSM) physics - we expect these would manifest in the form of TeV scale resonances. To make connection with theoretical models, we would need to measure both the mass of such a resonance and its coupling to the SM particles. However, it is not straightforward to measure the couplings of a massive particle in a hadron collider environment, where the momenta of the interacting partons are not known for any single event. Generally,  it is easier to measure ratios of couplings or production rate times branching ratios (BRs). From these, in principle, it is possible to measure any coupling if the total decay width of the particle is also known. However, since unlike the BRs, measuring the width of a particle accurately can be 
quite difficult, this way of measuring couplings may lead to large uncertainties. In this note we present a method of extracting couplings at a hadron collider that depends only on the BRs but not on the measured width or the production mechanisms of the particle in question.

Motivation for this method comes from the simple observation that the decay of an off-shell particle is sensitive to the coupling involved. This idea, in itself, is not new and has been used, for example, in studies about constraining the Higgs width~\cite{coup-sim} and also in Ref. \cite{Gopalakrishna:2013hua} where it was hinted that it can be used to measure new physics couplings. What we propose in this note is a systematic way of extracting unknown couplings using this observation by identifying a set of new physical variables that are sensitive to the coupling of interest.

\section{Off-Shell Analysis and Coupling Extraction}
\label{sec:method}

When a massive unstable particle decays to lighter particles, the invariant mass of the daughter particles shows a distribution peaked about the mass of the particle. The shape of the distribution is well approximated by the famous Breit-Wigner distribution around the peak (resonance) and its width gives us the decay rate. Discovery of a new unstable particle usually entails looking at this distribution, designing cuts to isolate regions close to the peak as its position gives us the mass of the particle. However, the same distribution can also be used to  extract information about the coupling involved in the decay since both the width and the height of the distribution depend on it. If instead of looking near the resonance we look at regions away from it, the cross-sections increase with the coupling as the width grows. At the same time, the height of the peak decreases with the increasing coupling keeping the total cross-section roughly the same. In other words, when the coupling involved 
in the decay increases, the distribution spreads out keeping the area covered roughly constant. This tells us it may be possible to learn about the coupling involved in the decay if instead of the total cross-section we focus on parts of the phase-space.

To begin with, let us consider the single production of a heavy particle $\Phi$ within a toy model at the LHC. The $\Phi$ decays to two SM particles, $q$ and $q'$, with the decay controlled by an unknown coupling $\lm_{qq'}$. The actual method of production of $\Phi$ does not concern us here as we are interested in its decay and the final state kinematics. Our analysis would hold even if $\Phi$ is produced in association with a different particle or produced in pairs. 

The width-dependent part of the amplitude for the process $p p\to\Phi\to qq'$ can be written as:
\ba
\mc M \lt(\Phi\to qq^\prime\rt)\propto \lm_{ qq'} /\lt[ \lt(p_\Phi^2 - M_{\Phi}^2\rt) + i \Gm_\Phi(\lm_{qq'}) M_{\Phi}\rt],
\ea
where $p_\Phi$ is the momentum of $\Phi$. From this, it is clear that for $p_\Phi^2 \approx M_{\Phi}^2$, i.e., near the resonance, the cross-section scales as $(\lm_{ qq'}/\Gm_\Phi)^2$. If $\Phi\to qq^\prime$ is the dominant decay mode of $\Phi$, then $\Gm_\Phi \sim \lm_{ qq'}^2$ and so $(\lm_{ qq'}/\Gm_\Phi)^2\approx 1/\lm_{ qq'}^2$. On the other hand, in the kinematic region where $|p^2-M_{\Phi}^2| \gg \Gm_{\Phi} M_{\Phi}$, the cross-section scales like $\lm_{qq'}^2$. 

Our minimal assumptions at this point are that both $M_{\Phi}$ and BR($\Phi\to qq'$) are known fairly accurately from experiments.\footnote{With these two assumptions, extracting the couplings involved in the decay of $\Phi$ is equivalent to measuring its decay width. Hence the recent studies about constraining the Higgs width~\cite{coup-sim} are actually similar in spirit to this study.}
Below, we outline the main steps that one could follow to obtain $\lm_{qq'}$.
\begin{enumerate}
\item
 The first step is to numerically simulate the process for many different values of $\lm_{qq'} = \lm_{qq'}^i$ using $\Gm^i_{\Phi}$ obtained from the theoretically computed partial width and the experimentally measured BR as $\Gm^i_{\Phi}=\Gm^i_{qq'}/\textrm{BR} (\Phi\to qq)$. This way, one can avoid using the measured width but rely on the BR that can be measured more accurately.

\item
 The regions \emph{outside} the resonance can be accessed with an invariant mass cut on $M(qq')$ defined by a new parameter $\gm_O$ that parametrizes the degree of off-shellness of the intermediate $\Ph$:
\ba
 |M(qq')-M_{\Phi}| &\geq \gm_{O} M_{\Phi}\, .\label{eq:phioff-cut}
\ea
We note that the above cut is similar to the one used in~\cite{Gopalakrishna:2013hua}.
For each $\lm_{qq'}^i$ and for various choices of $\gm_O$, one could compute the following ratio:
\ba 
 \mc R(\lm_{qq'}^i)\bigg|_{\gm_R,\gm_O} =\frac{\sg_{\Phi}(\lm_{qq'}^i,\gm_O)}{\hat{\sg}_{\Phi}(\lm_{qq'}^i,\gm_R) }\,,
\ea 
where $\hat{\sg}_{\Phi}$ is the \emph{on-shell} cross-section, computed using a cut that isolates the resonance:
\ba
|M(qq')-M_{\Phi}| &\leq \gm_{R} M_{\Phi}\;.\label{eq:phion-cut}
\ea
The advantage of working with the ratio $\mc R(\lm_{qq'}^{i})$ instead of either cross-section is that most uncertainties associated with the production mechanism involving pdfs etc. get canceled in the ratio. And more importantly, any other unknown coupling present in the production of $\Ph$, i.e., $pp \to \Ph$, also gets canceled. Now, for each combination of $\{\gm_R,\gm_O\}$, one can prepare a ``calibration curve'' (CC) by interpolating between $\mc R(\lm_{qq'}^i)$'s and thus prepare a family of such curves. 

\item
 From the experiment one can measure the ratio $\mc R = \mc R_{\rm exp} \pm \Delta \mc R_{\rm exp}$ for all the $\{\gm_R,\gm_O\}$ combinations considered to prepare the CCs. Here, we have accounted for errors in the cross-section measurements by a factor $\Delta \mc R_{\rm exp}$. Now, in principle, one could simply match $\mc R_{\rm exp} $ for any single combination of $\{\gm_R,\gm_O\}$ with the corresponding theoretical CC and read off the coupling. However, there are some difficulties associated with this procedure. The efficiency of coupling extraction is not uniform for a particular CC since it depends on the steepness (or slope) of the curve - for instance, if a particular CC is almost flat, it is not possible to extract a unique value of the coupling by matching on to the experiment.  \textit{A priori}, it is not clear how to select the optimal combination of $\{\gm_R,\gm_O\}$. 
One could also take the average of the couplings extracted from different CCs but the issue of assigning  proper weight factors for averaging may become an ambiguous issue. Instead, one can extract $\lm_{qq'}$ from a simultaneous fit using all the CCs. This can be done, e.g., by maximizing a likelihood function defined as:
\ba
L = \prod_k \frac{1}{\sqrt{2\pi}\;\Delta \mc R^{k}_{\rm exp}} \exp\left[-
\left(
\frac{\mc R^{k}\left(\lm_{qq'}\right) - \mc R^k_{\rm exp}}
       {\sqrt{2}\;\Delta \mc R^{k}_{\rm exp} }  
\right)^2
\right],\label{eq:likelihood}
\ea 
where the index $k$ runs over all the combinations of \{$\gm_R,\gm_O$\}. The $\lm_{qq'}$ thus extracted will correspond to the value for which $\mc R\left(\lm_{qq'}\right)$ best describes $\mc R_{\rm exp}$ for all the \{$\gm_R,\gm_O$\} combinations.
\end{enumerate}

For these outlined steps to work optimally, the process in question should have a large enough cross-section so that there are an appreciable number of events left over after the off-shell cuts. If the cross-section is small it may not be possible to measure $\mc R_{\rm exp}$ reasonably accurately for many \{$\gm_R,\gm_O$\} combinations which in turn will increase the error in fitting. Also, the method works better with a broad resonance. Thus far, our discussion has been purely about the signal process. However, in practice, handling the SM backgrounds is an important issue. The preparation of the CCs should thus also include the effect of the SM background and its interference with the signal. Inclusion of these effects could change the dependence of the CCs on the coupling (due to the interference term). However, since the method does not rely on the nature of the dependence and only on the fact that there is one, we expect the outlined steps to work. We demonstrate this within the context of an 
illustrative example in the next section.

\section{Illustrative Example}
\label{sec:toy}
To demonstrate the above method with an example, we consider a simple model with a new, heavy  color triplet $b^{\prime}$ quark with electric charge $-1/3$. \footnote{Such heavy fermions are part of many well-motivated extensions of the SM~\cite{Gopalakrishna:2013hua,vectorquarks}.} We make the simplifying assumption that it decays to $tW$ via: 
\begin{eqnarray}
\label{eq:lagran}
\mathcal{L}_{int} \supset
\frac{g_W}{\sqrt{2}}\lm_{tW} ~\bar{t}_{L}\gamma^{\mu}b^{\prime}_{L}W^{+}_{\mu} + \textrm{H.c.}
\end{eqnarray}
In the Lagrangian above, $\lm_{tW}$ is the unknown coupling that we want to extract.
At the LHC, if the $b^\pr$ is not too
heavy, QCD mediated pair production will be the dominant production channel. 
There will also be the $W$ mediated single production channel of $b^\prime$, like $qq'\to b't$. Following the steps described before one can use the single production process to probe $\lm_{tW}$. 
However, as this process is weak interaction mediated, its cross-section will be quite small and it will be more so if $\lm_{tW}$ is also small. Hence, to take advantage of the enhanced cross-sections, here we make use of pair production and simply apply off-shell cuts to one of the two final state particles - this reiterates our earlier point that this analysis depends only on the kinematics of the final state particles and we are free to choose the particular production process that would serve our purpose well. We note, however, that  in our analysis below we will include all possible contributions to the $pp\to b^{\pr}tW\to tWtW$ process, although the pair production will dominate.

\begin{figure}[t]\bc
\includegraphics[angle=0,origin=c,width=0.9\linewidth]{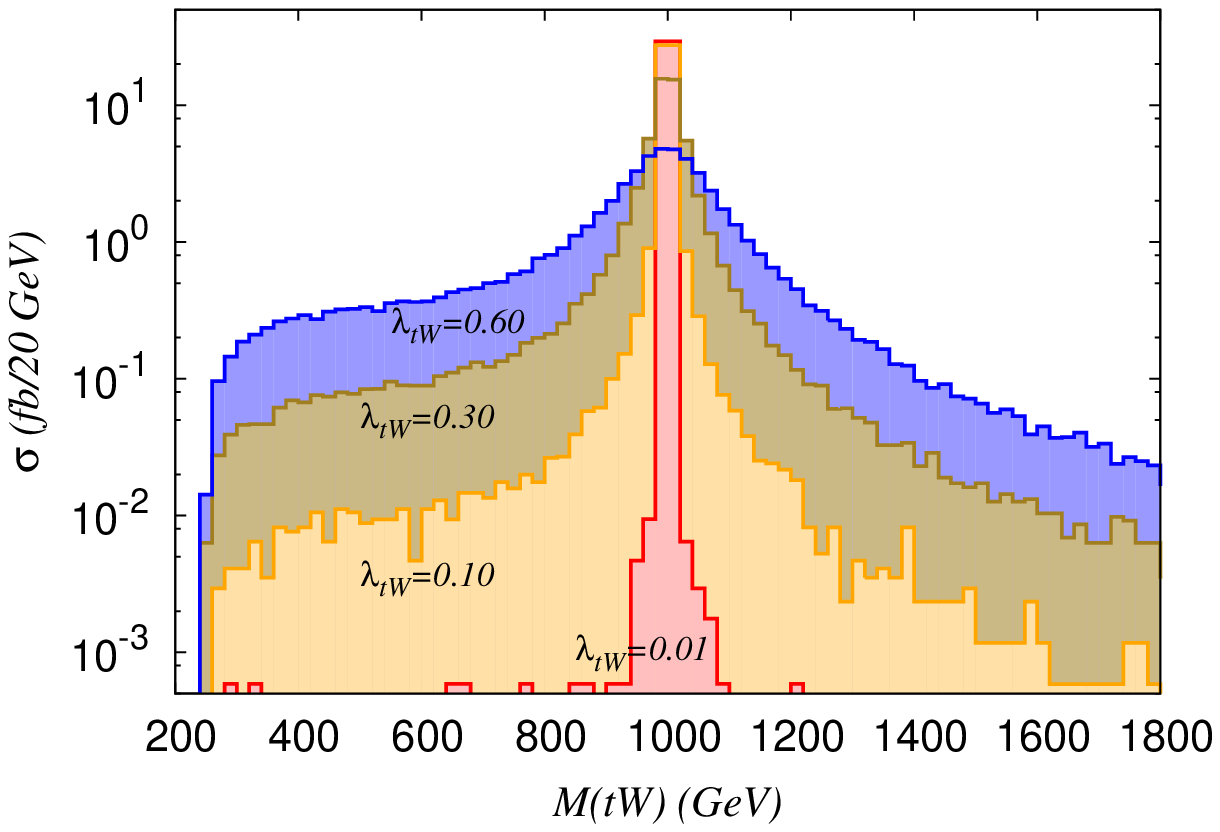}
\caption{\label{fig:dist} The invariant mass distribution of the $tW$ pair in the process $pp \to b^\pr tW$ for different values of $\lm_{tW}$ (events generated with with \textsc{Madgraph5} \cite{Alwall:2011uj}). Here, we fix $M_{b^\pr}=1$ TeV and BR$(b^\pr\to tW)=0.75$.}\ec
\end{figure}

In Fig. \ref{fig:dist} we show the invariant mass distributions of the $tW$ pair in $pp\to b^{\pr}tW$ for different values of $\lm_{tW}$. As $\Gm_{b^{\pr}}$ increases with increasing $\lm_{tW}$, the distribution spreads out and, with the increasing width, the height of the distribution decreases so that the area contained by the distribution remains approximately constant, i.e, the total cross-section is not very sensitive to $\lm_{tW}$, reiterating our point at the beginning of the last section.
But as can be seen, there \emph{are} parts of the phase space that are sensitive to $\lm_{tW}$. For example, if we look away from the mass peak, i.e., when the $tW$ pair is not coming from a close to on-shell $b^\pr$, the cross-section becomes highly sensitive to $\lm_{tW}$, in keeping with our discussion earlier.
Of course, the price to pay for looking away from the resonance is the reduced cross-section compared to the resonant production. However, we can mitigate this to a large extent by considering ``mixed states'' where one $b^\pr$ is produced resonantly.

To extract $\lm_{tW}$, we follow the steps outlined before. Assuming $M_{b^{\pr}}$ and ${\rm BR}(b^\pr\to tW)$ are known from experiments, we generate events for
$pp\to tWtW$ in \textsc{Madgraph5} \cite{Alwall:2011uj} (including both $b^\pr$ and SM contributions as well as their interference) for different values of $\lm_{tW}$ where we compute $\Gm_{b^\pr}^{i}$ as $\Gm_{b^{\pr}}^{i}=\Gm_{tW}^{i}/{\rm BR}(b^\pr\to tW)$. For this example, we set $M_{b'}=1$~TeV and ${\rm BR}(b^\pr\to tW)=0.75$ and generate events for the 14 TeV LHC. Since in this case we are considering pair production, rather than using the cut defined in Eq.~\ref{eq:phioff-cut}, we modify it to include two cuts on the $tW$ pairs, one on-shell and one off-shell:
\begin{subequations}\begin{align}
 \tr{ I. } \quad |M(t_1W_p)-M_{b^{\pr}}| &\leq \gm_{R} M_{b^{\pr}}\;,\label{eq:on-cut}\\
\tr{II. } \quad |M(t_2W_q)-M_{b^{\pr}}| &\geq \gm_{O} M_{b^{\pr}},\label{eq:off-cut}
\end{align}\end{subequations}
where $p,q=\{1,2\}$  or $\{2,1\}$ and the numbers imply that the particles are $p_{\rm T}$-ordered. Condition (I) reconstructs the resonant $b^\pr$ while  condition (II) accesses the $\lm_{tW}$ sensitive off-shell region.
Our expression for $\mc R(\lm_{tW}^{i})$ is given by:
\ba 
\mc R(\lm_{tW}^{i})\bigg|_{\gm_R,\gm_O} = \frac{\sg_{tWtW}(\lm_{tW}^{i},\gm_R,\gm_O)}{\hat{\sg}_{b^\pr b^\pr}(\lm_{tW}^{i},\gm_R) },\label{eq:bprR}
\ea
where $\hat{\sg}_{b^\pr b^\pr}(\lm_{tW}^{i},\gm_R)$ is now computed by applying cuts as in Eq.~\ref{eq:on-cut} to reconstruct two resonant $b'$s.
We produce a family of CCs for $\mc R$'s with various combinations of \{$\gm_R,\gm_O$\}. 
\begin{figure}[t]\bc
\includegraphics[angle=0,origin=c,width=0.9\linewidth]{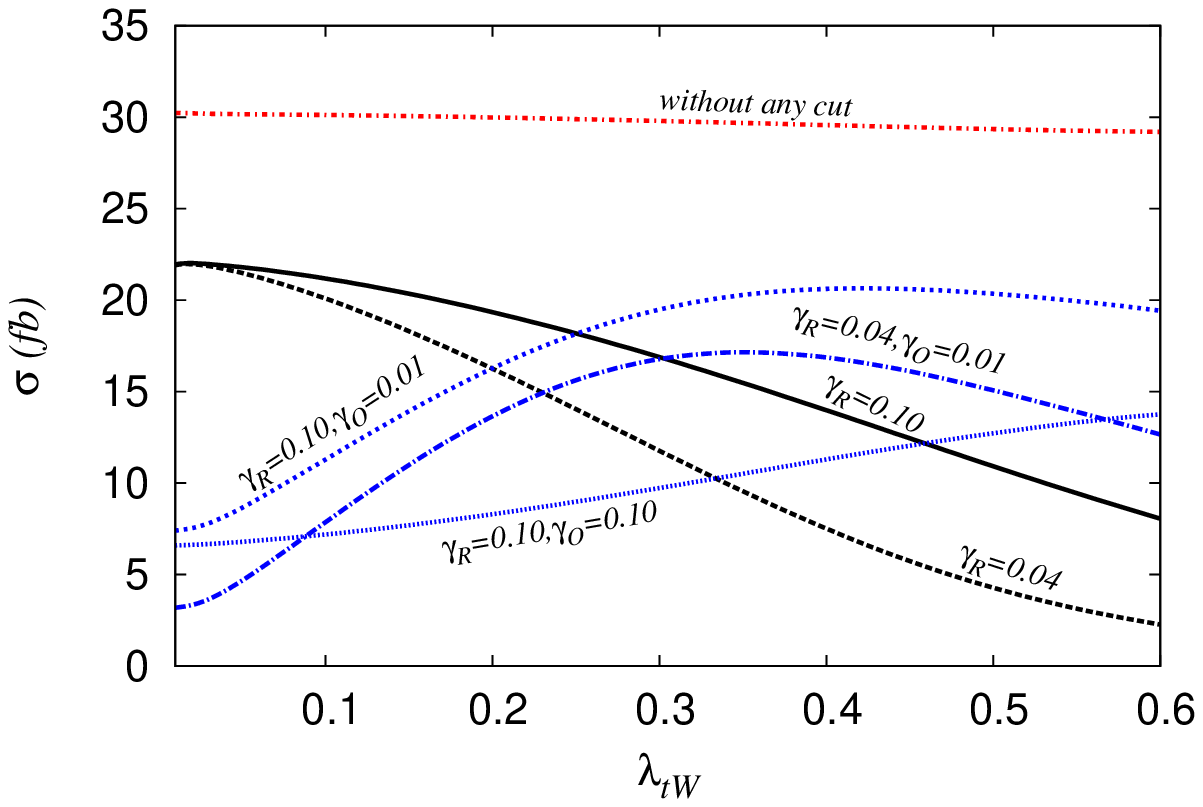}
\caption{\label{fig:cali} The dependence of $\sg_{tWtW}\left(\lm_{tW},\gm_R,\gm_O\right)$ on $\lm_{tW}$ with $\gm_{R/O}$ cuts defined in
Eqs.~(\ref{eq:on-cut}) \& (\ref{eq:off-cut}). The curves marked with only $\gm_R$ show $\hat\sg_{b'b'}\left(\lm_{tW},\gm_R\right)$.}\ec
\end{figure}
In Fig.~\ref{fig:cali} we plot $\sg_{tWtW}$ (including SM and BSM contributions)  with respect to $\lm_{tW}$
before and after cuts for three different choices of $\{\gm_R,\gm_O\}$. We also show $\hat\sg_{b'b'}$ in the same figure. The total cross-section $\sg_{tWtW}$ without any cut is mostly insensitive to $\lm_{tW}$ -- it remains almost constant for smaller $\lm_{tW}$, but slightly decreases in the higher $\lm_{tW}$ region (mostly) because of the larger widths. In this example the conditions \ref{eq:on-cut} and \ref{eq:off-cut} cuts become competitive in nature as they are applied on different $tW$ pairs simultaneously. Applying the cut in Eq.~\ref{eq:on-cut} we reconstruct the resonant $b^\pr$, i.e., we accept the $tW$ pairs that fall {\it inside} the $\gm_R$ mass window and so an increasing $\gm_{R}$ means accepting more events (larger cross-section). Whereas, with Eq.~\ref{eq:off-cut} we look away from the peak (we accept the $tW$ pairs that fall {\it outside} the $\gm_O$ mass window) and so an 
increasing $\gm_O$ results in reduced cross-section. With this in mind, it 
becomes 
easier to understand the curves with 
the $\{\gm_R,\gm_O\}$ cuts. For example, for a fixed $\gm_R$, we expect the cross-section to go down with increasing $\gm_O$. This we can see by comparing the curves with \{0.10, 0.01\} and \{0.10, 0.10\} -- the second one is smaller than the first for all values of $\lm_{tW}$. Similarly for a fixed $\gm_O$, if we decrease $\gm_R$ we should get a smaller cross-section which can be seen from the curves with \{0.10, 0.01\} and \{0.04, 0.01\}. We also observe that with increasing $\lm_{tW}$, the $\sg_{tWtW}$ curves first increase then decrease. 
This is because for a fixed $\gm_R$, as we increase $\lm_{tW}$ (and hence $\Gm_{b^\pr}$), we miss more and more $tW$ pairs that come from the resonant $b^\pr$, i.e., reconstruction efficiency of the resonant $b^\pr$ decreases and as a result the cross-section decreases. This also explains why the $\hat \sg_{b^\pr b^\pr}$ curves, for which we are reconstructing two resonant $b^\pr$'s, fall (and fall faster compared to $\sg_{tWtW}$ curves) with increasing $\lm_{tW}$. When the coupling $\lambda_{tW}\to$ 0, the difference between the $\gamma_R$ curves and the total cross-section is accounted for by the SM background.

To see how the method performs we have considered four different test couplings $\lm_{tW} = \lm_{tW}^{\rm test}$ and use $\mc R(\lm_{tW}^{\rm test})$ in place of $\mc R_{\rm exp}$ (Eq. \ref{eq:bprR}), assuming all $\mc R(\lm_{tW}^{\rm test})$'s have uniform 10\% errors for all combinations of $\{\gm_R,\gm_O\}$.  We have considered eight different values of $\gm_{R}$ between  $0.02$ and $0.16$ and ten different values of $\gm_O$ between $0.01$ and $0.10$.  We present the results of our analysis in Table \ref{tab:fit}.  We see that we are able to extract the couplings with percent-level accuracies. We note, however, that ours is a parton level analysis. Doing full detector simulations and considering SM backgrounds properly are likely to modify the extraction efficiency. These issues are currently under investigation \cite{work}.
\begin{table}[t]
\centering
\begin{tabular}{ccc}
\hline 
$\lm_{tW}^{\rm test}$ &$\quad$& $\lm_{tW}$ obtained\\ 
\hline \hline
0.08125 &$\quad$& 0.083 $\pm$ 0.0020  \\ 
0.17338 &$\quad$& 0.171 $\pm$ 0.0020   \\ 
0.32785 &$\quad$& 0.328 $\pm$ 0.0021  \\ 
0.43597 &$\quad$& 0.435 $\pm$ 0.0026  \\ 
\hline 
\end{tabular}
\caption{\label{tab:fit} The different choices for $\lm^{\rm test}_{tW}$ and the corresponding values obtained by maximizing the likelihood function defined in Eq. \ref{eq:likelihood}. The errors in the fitted values correspond to $1\sg$.} 
\end{table}
\section{Discussions}
\label{sec:discussions}

In this note, we presented a novel method of extracting the couplings of new, heavy states to SM particles. The essential point underlying the method is a rather simple observation that when a particle decays inside a collider, the invariant mass distribution of its decay products retains information of the coupling involved in the decay. 
With the off-shell cuts, this coupling extraction procedure actually becomes sensitive to the shape of the invariant mass distribution.
The method is largely free of modeling assumptions apart from the fact it uses theoretically computed partial decay width to avoid the use of the measured total width of the new particle, and thus avoids the errors associated with width measurement. Of course, computing the total width using the theoretically computed partial width and the experimentally measured BR means that some errors could come from these sources and ideally one should account for both. However, since generally these errors are expected to be much less severe than the ones in measuring widths experimentally, we neglect these for the time being. Similarly we neglect any error in measuring the mass of the new particle.
The method does not depend on the exact production mechanism of the new particle.
The coupling extraction method outlined here is completely general and applicable to any new fermionic or bosonic state. It also has the advantage of being independent of collider details, or the actual values of the mass and the BRs of the particle in question. 

In this note, we have ignored the issue of considering the complete SM backgrounds which in reality is a very crucial one. Since realistic background can only be considered in a case-by-case basis, inclusion of simple parton level background in the example shown should only be considered as an illustration. However, we hope that after the discovery of a new particle, its background will also be fairly well known and thus can be dealt with.  Since our aim here is simply to outline the methodology, we postpone a more complete demonstration with detector level simulations for both signal and background to a future publication \cite{work}.

To extract any coupling the cross-section of the process in question needs to be large enough to start with so we have enough events left after all the 
cuts. 
It is interesting to contrast this method with the usual discovery procedure. If the new state has a significant width (comparable to its mass), discovery is rendered more difficult as we would not have the typical sharp peak. On the other hand, a large width guarantees that regions away from resonance do not have negligibly small cross-sections and hence coupling extraction becomes more feasible.

Of course, one need not restrict attention to new physics alone. In principle, this method can be applied to extract the SM CKM matrix element $V_{tb}$ from
top pair production. In fact, it can be applied to extract any coupling, be it SM or BSM, as long as enough events remain after the cuts. 

\begin{acknowledgements}
We thank Heather Logan and Biswarup Mukhopadhyaya for helpful comments on the draft, Shrihari Gopalakrishna and Erich Varnes for useful discussions. BC is supported by  the Department of Energy under Grant~DE-
FG02-13ER41976.  TM is partially supported by funding from the DAE, for the RECAPP, HRI. SM acknowledges financial support from
the CNRS. BC acknowledges the hospitality of IMSc where this project was started.

\end{acknowledgements}

\end{document}